\documentclass[aps,nofootinbib,superscriptaddress,twocolumn,prd,10pt]{revtex4-1}

\usepackage{graphicx}
\usepackage{amsmath,amssymb}
\usepackage{hyperref}
\usepackage{braket}
\usepackage{booktabs}
\usepackage{subfigure}
\usepackage{float}
\usepackage[dvipsnames]{xcolor}
\usepackage{mathrsfs}
\usepackage{comment}
\usepackage{bm}
\usepackage[normalem]{ulem}

\hypersetup{
    colorlinks=true,
    linkcolor=red,
    citecolor=blue,
}

\newcommand{\dcal}{\ensuremath{\delta_\mathrm{cal}}}
\newcommand{\rmx}{\mathrm{X}}
\newcommand{\white}{\mathrm{w}}
\newcommand{\clustered}{\mathrm{c}}
\newcommand{\poisson}{\mathrm{p}}
\newcommand{\radio}{\mathrm{r}}
\newcommand{\dust}{\mathrm{d}}
\newcommand{\sync}{\mathrm{s}}
\newcommand{\ame}{\mathrm{a}}
\newcommand{\GHz}{\mathrm{GHz}}
\newcommand{\cobefiras}{\textit{COBE}/FIRAS}
\DeclareMathOperator{\csch}{csch}

\begin{document}

\title{Measuring $\mu$-Distortions from the Thermal Sunyaev-Zeldovich effect}

\author{David Zegeye}
\affiliation{Kavli Institute for Cosmological Physics, Department of Astronomy \& Astrophysics, Enrico Fermi Institute, The University of Chicago, Chicago, IL 60637, USA}
\affiliation{Department of Astronomy \& Astrophysics, The University of Chicago, Chicago, IL 60637, USA}

\author{Thomas Crawford}
\affiliation{Kavli Institute for Cosmological Physics, Department of Astronomy \& Astrophysics, Enrico Fermi Institute, The University of Chicago, Chicago, IL 60637, USA}
\affiliation{Department of Astronomy \& Astrophysics, The University of Chicago, Chicago, IL 60637, USA}

\author{Wayne Hu}
\affiliation{Kavli Institute for Cosmological Physics, Department of Astronomy \& Astrophysics, Enrico Fermi Institute, The University of Chicago, Chicago, IL 60637, USA}

\date{\today}

\begin{abstract}
The thermal Sunyaev-Zel’dovich (tSZ) effect is a spectral distortion of the cosmic microwave background (CMB) resulting from inverse Compton scattering of CMB photons with electrons in the medium of galaxy clusters. The spectrum of the tSZ effect is typically calculated assuming the spectrum of the CMB is a blackbody. However, energy or photon number injection at any epoch after photon creation processes become inefficient will distort the blackbody, potentially leading to a chemical potential or $\mu$-distortion for early injection.   These \textit{primordial} spectral distortions will therefore introduce a change in the tSZ effect, effectively a distortion of a distortion. While this effect is small for an individual cluster's spectrum, upcoming and proposed CMB surveys expect to detect tens of thousands of clusters with the tSZ effect. In this paper, we forecast constraints on the $\mu$-distortion monopole from the distortion of the tSZ spectrum of clusters measured by CMB surveys.
We find that planned experiments have the raw sensitivity to place  constraints on $\mu$ that are comparable to or better than existing constraints but control over foregrounds and other systematics will be critical.
\end{abstract}

\maketitle
\section{Introduction}
\label{sec:introduction} 

The cosmic microwave background (CMB) provides us with vital information about the origin and evolution of our observable universe, and of the underlying physical laws that govern it. We have greatly improved measurements of CMB temperature and polarization anisotropy over the last 20 years with experiments such as \textit{WMAP} \cite{WMAP:2012fli} and \textit{Planck} \cite{Planck:2018nkj}. 

On the other hand, our constraints on the frequency spectrum of the CMB have not improved since the measurements of the Far Infrared Absolute Spectrophotometer on the \textit{Cosmic Background Explorer} (\cobefiras\ hereafter) \cite{Fixsen:1996nj}. Although the measured CMB spectrum closely matches a blackbody, the CMB in fact is expected to have some small distortion away from a blackbody spectrum.

Energy injections in the form of diffusion damping of small-scale anisotropies, resulting from imperfect photon-baryon coupling in the pre-recombination plasma, during periods of inefficient thermalization ($z \lesssim 2 \times 10^6$) will slightly distort the spectrum. For $2\times 10^6 \gtrsim z \gtrsim 5 \times 10^4$, distortions of the $\mu$-type are generated by this process, while for $5\times10^4 \gtrsim z \gtrsim 1100$ distortions of the $y$-type are generated. Using an internal blackbody as a calibrator, \cobefiras\ was able to confirm the CMB spectrum closely follows a blackbody distribution and place upper limits of $|y| < 1.5 \times 10^{-5}$ and $|\mu| < 9 \times 10^{-5}$ (95\% CL).

While diffusion damping of fluctuations from slow-roll inflation is one small 
($\mu \gtrsim 10^{-8}$)
but guaranteed method for generating spectral distortions before recombination
\cite{1970Ap&SS...7....3S,1991ApJ...371...14D,Hu:1994bz,Chluba:2012gq}, other possibilities
include annihilating particles \cite{Bolliet:2020ofj}, diffusion damping in  inflationary models that generate primordial black holes \cite{Zegeye:2021yml}, primordial black hole evaporation \cite{Acharya:2019xla}, and primordial gravitational waves \cite{Ota:2014hha}.

The CMB radiation can also be distorted by post-recombination sources, for example through inverse Compton scattering off of the  hot electron gas in galaxy clusters, resulting in cluster-scale distortions of the CMB spectrum, {a phenomenon  referred to as the thermal Sunyaev-Zel'dovich (tSZ) effect \cite{1970Ap&SS...7....3S}}.
In general spectral distortions allow one to
probe any process associated with energy injection into the CMB after the thermalization epoch.

Measuring the mean, or monopole, frequency spectrum of the CMB is extremely challenging, because it requires an experiment to retain information about the absolute power received from the sky, not just the difference in power between different sky locations. Absolute measurements require exquisite stability over long timescales and tight control over any spatially varying sources of emission. Note that these stability requirements remain even for an experiment that does not require an overall absolute gain calibration (e.g., \cite{Mukherjee:2018fxd,Mukherjee:2019pcq}). For these reasons, it is often assumed that such measurements can only be made from space.

One way around these requirements is to measure the mean distortion of an anisotropic signal that can be measured differentially, such as the CMB dipole \cite{Balashev:2015lla} or primary CMB anisotropy. The issue with using CMB temperature anisotropy is that most differential CMB experiments use the temperature anisotropy (either the dipole---or, more precisely, the annual modulation of the dipole---or the degree-scale and smaller anisotropy) as a calibration source, with the underlying assumption that the photon distribution follows a perfect blackbody. This effectively destroys any sensitivity to spectral distortions from the dipole or primary anisotropy, because the calibrated spectrum of the anisotropy will be forced to look like the derivative of a blackbody. Put another way, experiments designed to measure spectral distortions in the dipole or primary CMB anisotropy must find a different way of calibrating the relative response between observing frequencies.

In this work, we investigate the prospect for using the tSZ effect to measure monopole spectral distortions.
This method, first proposed by \cite{1980ApJ...241..858R}, was used recently by \cite{Luzzi:2021usi} to forecast constraints on the (primordial) $y$-distortion of the CMB from distortions of the (local-universe) tSZ effect. As discussed in \cite{DeZotti:2015awh}, this technique can in principle be applied to $y$- or $\mu$-type distortions, and was also proposed in \cite{1983IAUS..104..113W} to test the validity of early measurements indicating large spectral distortions near the blackbody peak, later demonstrated by \cobefiras\ to be spurious. Similar works have explored constraining the primordial recombination radiation \cite{Kholupenko:2014xna} and the redshift evolution of the CMB temperature from the distortion of the tSZ effect. This paper focuses on the potential constraints on the mean value of $\mu$-type distortions from measurements of the tSZ effect in the direction of massive clusters of galaxies using calibration from primary CMB temperature anisotropy under the blackbody assumption. We will forecast constraints on 
 this quantity 
from the upcoming CMB-S4 experiment \cite{Abazajian:2019eic} as well as one based on 
the proposed CMB-HD experiment \cite{CMB-HD:2022bsz}.

\section{CMB Spectral Distortions}

\subsection{$\mu$ and $y$ Distortions} 
\label{sec:mu}

At early epochs,  any changes in the photon phase space distribution $f$ are efficiently thermalized to a blackbody distribution through the joint action of the photon-number-changing processes double Compton scattering and Bremsstrahlung, and the energy-exchanging process (single) Compton scattering.  Number-changing processes fall out of equilibrium at  a redshift $z_i \sim 2 \times 10^6$ after which the 
photon distribution evolves mainly under the Kompaneets equation \cite{1957JETP....4..730K} (see Appendix~\ref{sec:relativistic} for relativistic corrections)
\begin{equation}
 \frac{\partial f}{\partial \tau} =  
 \frac{k_B T_e}{m_e c^2} \frac{1}{x_e^{2}}\frac{\partial}{\partial x_e} 
 \left[ x_e^4 
\left(\frac{\partial f}{\partial x_e} + f(1+f) \right) \right] ,
\label{eq:Kompaneets}
\end{equation}
where $\tau$ is the Thomson optical depth, and
  $x_e = h\nu/k_B T_e$ for a thermal distribution of electrons at temperature $T_e$. 
  The equilibrium distribution under the Kompaneets equation is a Bose-Einstein distribution. Any changes  to the number or energy density of the photons thereafter lead to a $\mu$-type distortion
\begin{equation}
\label{spectrum}
f=\left(e^{x+\mu}-1\right)^{-1},
\end{equation}
where $x=h\nu/k_B T$
with the temperature of the photons $T=T_e$.
For example a fractional energy injection of $\Delta\rho/\rho$ to the photons leads to $\mu \sim 1.4 \Delta \rho/\rho$.
Energy exchange via Compton scattering falls out of equilibrium 
at around $z_f \sim 5\times10^4$.  After this epoch, we can solve the Kompaneets equation by  plugging in the unperturbed spectrum (\ref{spectrum}) into the right hand side of Eq.~(\ref{eq:Kompaneets}) and integrating \cite{DeZotti:2015awh}
\begin{align}
    \Delta f(x, \mu, y) = \int d\tau \frac{\partial f}{\partial \tau} \approx y x e^{x + \mu}f^2  g(x,\mu),
    \label{eq:Deltaf}
\end{align}
with
\begin{align}
g(x,\mu) = 
    x {\coth \left(\frac{x + \mu}{2}\right)}-4, 
    \label{eq:g}
\end{align}
where  the Comptonization parameter,
\begin{equation}
y= \int d\tau \frac{k_B (T_e-T)}{m_e c^2},
\label{eq:y}
\end{equation}
is assumed to be $|y| \ll 1$.
This generalizes the standard expression for the $y$-type distortion to the case where $\mu\ne 0$, i.e.\ the photons possess an initial $\mu$-type distortion.
Notice that the spectrum only changes when  $T_e\ne T$, e.g.\ when the electrons are heated after $z_f$.
In particular we are interested in the case where 
the hot electrons exist in galaxy clusters and produce the late-time $y$-type distortions known as the tSZ effect.  Our generalization implies that in principle the initial $\mu$ value can be determined from a precise measurement of the tSZ spectrum.

\subsection{Interfrequency Calibration}
\label{sec:calibration}

As discussed in \S\ref{sec:introduction}, most differential CMB experiments derive their interfrequency calibration from CMB anisotropy, either the CMB dipole or the primary temperature anisotropy, under the assumption that the background photon distribution is a pure blackbody. Experiments that have access to very large angular scales, such as the {\it Planck} and {\it WMAP} satellites, calibrate off of the annual modulation of the dipole from the Earth's motion around the Sun. When compared to predictions using our precise knowledge of the current CMB temperature $T_0$ and the Earth's orbital velocity, and assuming a blackbody background, this provides both an interfrequency calibration and a calibration of the overall intensity scale. Experiments that use the primary anisotropy for interfrequency calibration (as is the case for most ground-based CMB experiments) need a separate reference for the absolute intensity scale, but since the inference for $\mu$ depends on the relative frequency dependence for a given amplitude $y$, an accurate relative calibration of channels is more important than the overall absolute calibration. For the specific measurement envisioned in this work, the absolute scale is effectively marginalized over, and we neglect it hereafter.

In practice, for the case of calibration off of the annual modulation of the dipole, the signal in each frequency band is scaled to agree with predictions assuming a pure blackbody background. 
The situation is similar for calibration off of the primary anisotropy: maps at every observing frequency $\nu$ are compared to each other in a region of the sky and a range of angular scales in which the primary CMB anisotropy is the dominant signal, and the maps are calibrated so that the signal follows the expected spectrum of temperature fluctuations in a background blackbody with mean temperature $T_0$.
In both cases, the true spectrum of the calibration source is that of temperature fluctuations in the true background, and the result of calibrating assuming a blackbody background is that the measured, calibrated dipole and/or primary CMB anisotropy is forced to follow the spectrum of temperature fluctuations in a blackbody.

Let us examine the case of calibrating off of the observed dipole in the presence of a monopole $\mu$ distortion in the background spectrum, while assuming the background spectrum is a blackbody.
(The results in the case of calibrating off of the primary CMB anisotropy are identical.)  
In the case of dipole calibration, the Lorentz invariance of $f$ implies that the specific intensity in the boosted frame $I^d_\nu \propto \nu^3 f$ obeys
\begin{equation}
I_\nu^d \propto  \frac{\nu^3} {e^{h\nu_{\rm rest}/k_B T +\mu}-1}
\end{equation}
where 
\begin{equation}
\nu_{\rm rest} = \left( \frac{1-\beta\cos\theta}{\sqrt{1-\beta^2}} \right) \nu,
\end{equation}
and $\theta$ is the angle between the line of sight and the velocity.
Notice that we can absorb the Doppler shift into a temperature anisotropy as usual and to first order in $\beta$, $T(\theta) \approx T(1+\beta\cos\theta)$. The change in the specific intensity becomes
\begin{equation}
\Delta I_\nu^d \approx ( \beta\cos\theta) \, T \frac{\partial I_\nu}{\partial T}.
\end{equation}
The frequency dependence involves  the derivative of $I_\nu$, and this result holds for calibration involving any type of temperature anisotropy by suitably generalizing the anisotropy source, not just a dipole due to a boost. 
Note that we are ignoring higher-order terms in the expansion of the blackbody fluctuation spectrum, which are negligible at least for the  order $10^{-5}$ anisotropy in the CMB.

If a blackbody background distribution is assumed in the calibration process, then the anisotropy-calibrated specific intensity $I_\nu^c$ differs from the true specific intensity $I_\nu$ by
\begin{equation}
I_\nu^c = C(x,\mu) I_\nu,
\end{equation}
where the miscalibration from the true spectrum is characterized as
\begin{equation}
C(x,\mu) = \frac{ \partial B_\nu  / \partial T} { \partial I_\nu/ \partial T}.
\label{eq:miscalibration}
\end{equation}

Notice that this anisotropy calibration factor involves the spectral shape of the derivative of the specific intensity not the specific intensity itself.  
Thus, while this particular calibration procedure removes any information about spectral distortions from the primary anisotropy signal, distortions of signals that do not have the spectrum of the 
temperature derivative of the CMB monopole spectrum can still be measured.

Counterintuitively, this observability includes the $\mu$-distortion of the CMB monopole itself:
\begin{align}
\frac{I_\nu^c(\mu)}{B_\nu} &= \frac{I_\nu}{B_\nu} \frac{\partial B_\nu/\partial T}{\partial I_\nu/\partial T}
= e^{-\mu} \frac{e^{x+\mu}-1}{e^x-1},
\end{align}
 and the correction for $|\mu|\ll x \ll 1$ goes as $\mu/x$.
In practice, as discussed in \S\ref{sec:introduction}, since this measurement requires a non-differential measurement on the sky, it remains challenging from the ground. 

\begin{figure}[t!]
\begin{center}
\includegraphics[width=1\columnwidth]{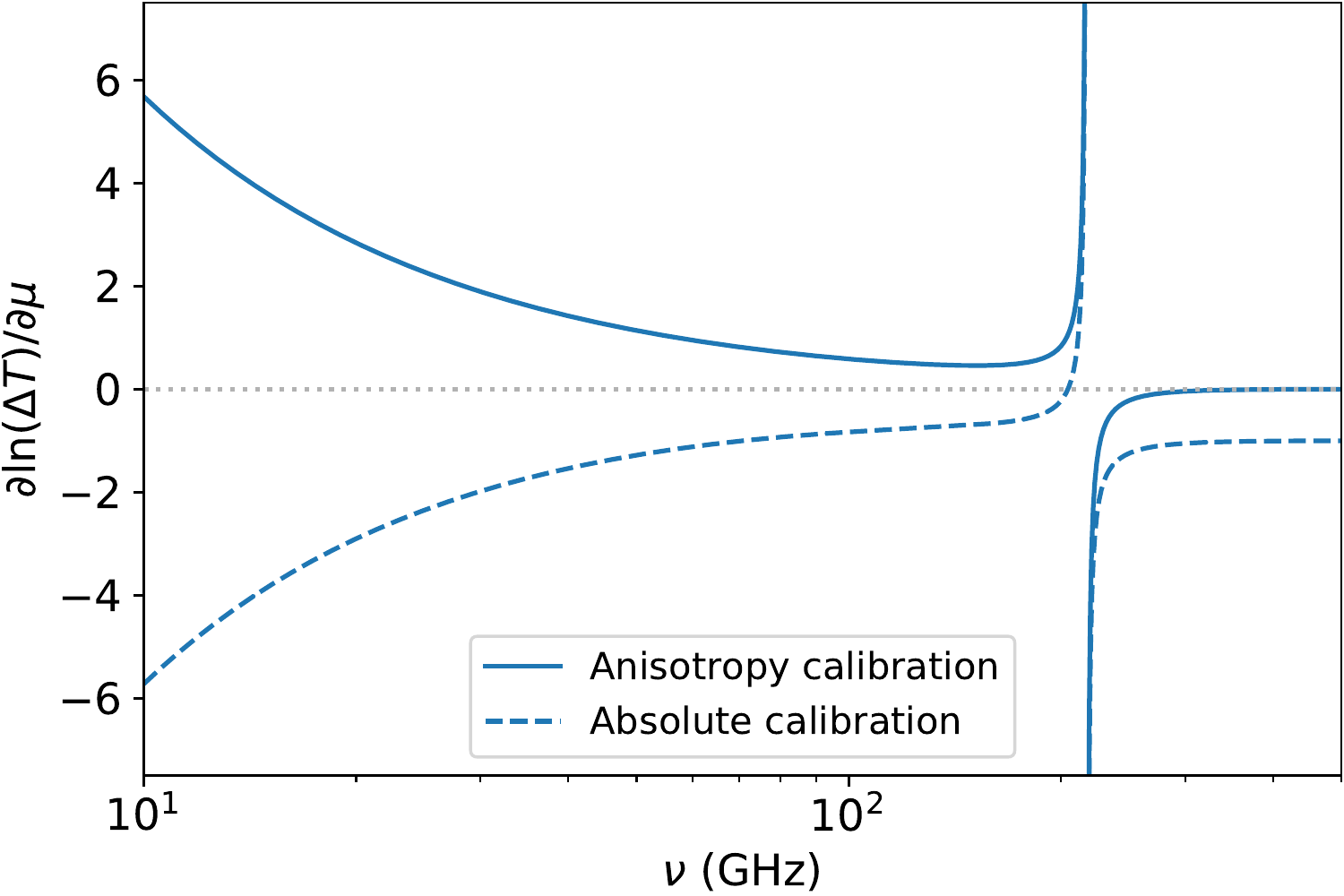}
\end{center}
\setlength\abovecaptionskip{0\baselineskip}
\caption{The fractional response of the tSZ temperature spectrum $\partial \ln  \Delta T/\partial \mu$ to a monopole $\mu$ distortion with anisotropy calibration as we assume in this work (solid blue line $\partial \ln g/\partial\mu$) vs.~absolute calibration (dashed blue line $\partial \ln (g/C) /\partial \mu$). The  dotted gray line denotes an undistorted spectrum for reference and the spike in the curves occurs at the tSZ null where the {\it fractional} response diverges corresponding to a finite change in the location of the null. 
\label{fig:muy}}
\end{figure}

Now let us apply this sort of calibration to the tSZ distortion of a $\mu$-distorted background in the direction of a  galaxy cluster, a signal which can be measured differentially.  
In terms of the calibrated apparent CMB temperature fluctuation at frequency $\nu$, $\Delta T$, we obtain
\begin{align}
\Delta T(x,\mu) & \equiv \frac{\Delta I_\nu^c}{\partial B_\nu/\partial T} = \frac{\Delta I_\nu}{\partial I_\nu/\partial T} 
= \frac{\Delta f}{\partial f/\partial T} \nonumber\\
& = y T_0 g(x,\mu),
\label{eq:calibratedSZ}
\end{align}
where we have used Eq.~(\ref{eq:Deltaf}) for $\Delta f$.
Notice that the anisotropy calibrated $\Delta T$ differs from the absolutely calibrated  temperature fluctuation, and Eq.~(\ref{eq:g}) for $g(x,\mu)$ carries the measurable  frequency dependence under anisotropy calibration.    This difference is illustrated in 
Fig.~\ref{fig:muy}.  Notice also that in both cases the response to $\mu$ increases at low frequency but with the opposite sign.

\subsection{Cluster model}

Using Eq.~(\ref{eq:y}), our model for the value of the Compton $y$ parameter in the direction of an isothermal cluster ($T_e = {\rm const.} \gg T$) is
\begin{equation}
 y(\theta) = \frac{k_{B} T_{e}}{m_{e} c^{2}} 
\tau(\theta),
\label{eq:compton}
\end{equation}
where $\theta$ is the angular distance from the center of the cluster.
For the optical depth profile $\tau(\theta)$ we follow e.g., \cite{SPT:2014wbo}, and adopt a spherically symmetric $\beta$ model or King profile with $\beta=1$ 
and express Eq. (\ref{eq:compton}) as
\begin{equation}
    y(\theta)=y_c\left[1+\left(\frac{\theta}{\theta_{\mathrm{c}}}\right)^{2}\right]^{-1}.
    \label{eq:ytheta}
\end{equation}
Here, the angular size of the cluster's core is given by $\theta_{c}=r_{c} / D_{A}$, with $D_A$ being the 
angular diameter distance and $r_c$ the core radius of the cluster, all in comoving coordinates for later convenience. 
We follow \cite{Plagge:2009eu, Liu:2014pqa} and we adopt the relation $r_c \sim 0.2 R_{500c}$,  where $R_{500c}$ is the radius
at which the
enclosed spherically averaged density is 500 times the critical density $\rho_{c}(z) \equiv 3 H^{2}(z) / 8 \pi {G}$.

 For $y_c$ we adopt the self-similar scaling relation
 \begin{equation}
y_c=A \tilde E^2(z) \left(\frac{M_{500 \mathrm{c}}}
{ 10^{14} M_{\odot} } \right)
\end{equation}
where 
\begin{equation}
\tilde E(z) \equiv \frac{H(z)}{70{\rm km/s/Mpc}},
\end{equation}
and the normalization $A$ from  X-ray cluster observations of luminosity and temperature at low $z$ \cite{Arnaud:2009tt} to calibrate  the universal pressure profile (Eq.~6 in Ref.~\cite{Hasselfield:2013wf})
\begin{equation}
A = 0.97 \times 10^{-5} h^{-3/2} .
\end{equation}
Note using this normalization in the context of Eq.~(\ref{eq:ytheta})
is approximate given differences with the universal pressure profile
\cite{Nagai:2007mt}.
We also adopt the temperature-mass relation \cite{Arnaud:2005ur}
\begin{equation}
k_B T_e  =2.28  \left(\frac{ M_{500c}}{ 10^{14} M_\odot }\tilde E(z) \right)^{0.585}  \text{keV} . 
\label{eq:temp}
\end{equation}

Because the noise in our forecasted surveys is expected to be diagonal in spherical harmonic ($\ell,m$) space, we choose to work in that basis. To transform Eq.~(\ref{eq:ytheta}) into $\ell,m$ space, we note that since even the most massive and low-redshift clusters only subtend a small angle on the sky, we can use the flat-sky approximation. As detailed in Appendix \ref{sec:flat}, in coordinates centered on the cluster at $\theta=0$, the spherical harmonic-space cluster profile is given by
\begin{equation}
y_{\ell m} =\sqrt{\frac{2\ell+1}{4\pi}} \delta_{m,0} y(\ell),
\end{equation} 
where
\begin{equation}
   y(\ell)=y_c 2 \pi \theta_c^2 K_0\left(  \ell  \theta_{\mathrm{c}}  \right),
   \label{eq:yell}
\end{equation} 
and $K_n(x)$ is the modified Bessel function of the second kind.

\section{Forecast}
\label{sec:forecast}

\subsection{Survey Specifications}

We forecast our constraints on $\mu$ from  tSZ cluster measurements using instrument configurations based on the upcoming CMB-S4 experiment and the proposed CMB-HD experiment. 
CMB-S4 will 
conduct two surveys: the Wide Survey conducted from Chile will cover 67\% of the sky, while
the Deep Survey will concentrate a similar amount of total observing weight on 3\% of the sky from the South Pole. From here on, we will refer to these two CMB-S4 surveys as ``S4-Wide'' and ``S4-Deep,'' respectively.
Both surveys will have similar beam sizes and differ mainly in the noise in the sky maps. The CMB-HD-like survey we forecast for here covers 50\% of the sky. For all three surveys, we use the instrument configuration parameters from Tab.~1 of \cite{Raghunathan:2021tdc}, which we reproduce in Tab.~\ref{tab:cmbsurvey}.
We note that for both S4-Wide and CMB-HD, the galactic plane will significantly contaminate our maps of tSZ clusters and reduce our ability to accurately measure the cluster spectrum. Therefore, for these surveys we assume $f_\text{sky} = 0.5$.

\begin{table}[]
\def\arraystretch{1.5}
\setlength{\tabcolsep}{3.75pt}
\begin{tabular}{|l|c|cccccc|}
\hline
\multicolumn{2}{|c|}{Channels (GHz)}                           & 30   & 40    & 90    & 150   & 220 & 270        \\\hline \cline{1-8} 
\multicolumn{1}{|c|}{Survey}           &          $f_\text{sky}$      & \multicolumn{6}{c|}{$\theta_\text{FWHM}$ \& $\sqrt{C_{\white}}$ ($\mu \mathrm{K}$-arcmin)} \\ \hline
S4-Wide                          & 50\%           & 7.3$'$  & 5.5$'$   & 2.3$'$   & 1.5$'$   & 1.0$'$ & 0.8$'$     
\\ &
& 21.8    & 12.4       & 2.0        & 2.0        & 6.9        & 16.7      \\ \hline
S4-Deep                          & 3\%            & 8.4$'$  & 5.8$'$   & 2.5$'$   & 1.6$'$   & 1.1$'$ & 1.0$'$        
\\ &
& 4.6  & 2.94       & 0.45       & 0.41       & 1.29       & 3.07      \\ \hline
CMB-HD                           & 50\%           & 1.4$'$  & 1.05$'$  & 0.45$'$  & 0.25$'$  & 0.2$'$ & 0.15$'$   
\\ &
 & 6.5    & 3.4        & 0.73       & 0.79       & 2.0        & 2.7       \\ \hline
\end{tabular}
\caption{Specifications for the CMB-S4 Wide and Deep surveys and a CMB-HD-like survey, taken from \cite{Raghunathan:2021tdc}.}
\label{tab:cmbsurvey}
\end{table}

\subsection{Cluster Catalog}

In addition to specifications on map noise,  angular resolution, and sky fraction, 
to forecast constraints on $\mu$ from the distortion of the tSZ spectrum we also need to define a sample of galaxy clusters. For each of the three surveys considered here, we use the expected cluster catalog for that survey, based on work from \cite{Raghunathan:2021tdc,Raghunathan:2021zfi}.

Underlying the expected  number of clusters detected by a given CMB experiment is the halo mass function $dn/d\ln M$, the number density of host dark matter halos at a given redshift $z$ over a logarithmic mass interval $d\ln M$. We adopt for this quantity the Tinker mass function \cite{Tinker:2008ff} as implemented in the publicly available code \texttt{Colossus}\footnote{http://www.benediktdiemer.com/code/colossus/} \cite{Diemer:2017bwl}.
Our cosmological parameters are taken from \textit{Planck} 2018 \cite{Planck:2018nkj}, where $\Omega_m = 1-\Omega_{\Lambda}=  0.31, \Omega_b = 0.049, H_0 = 67.7 \mathrm{km/s/Mpc}, \sigma_8 = 0.81, \tau = 0.054$, and $n_s = 0.965$ 

\begin{figure}[t!]
\begin{center}
\includegraphics[width=1\columnwidth]{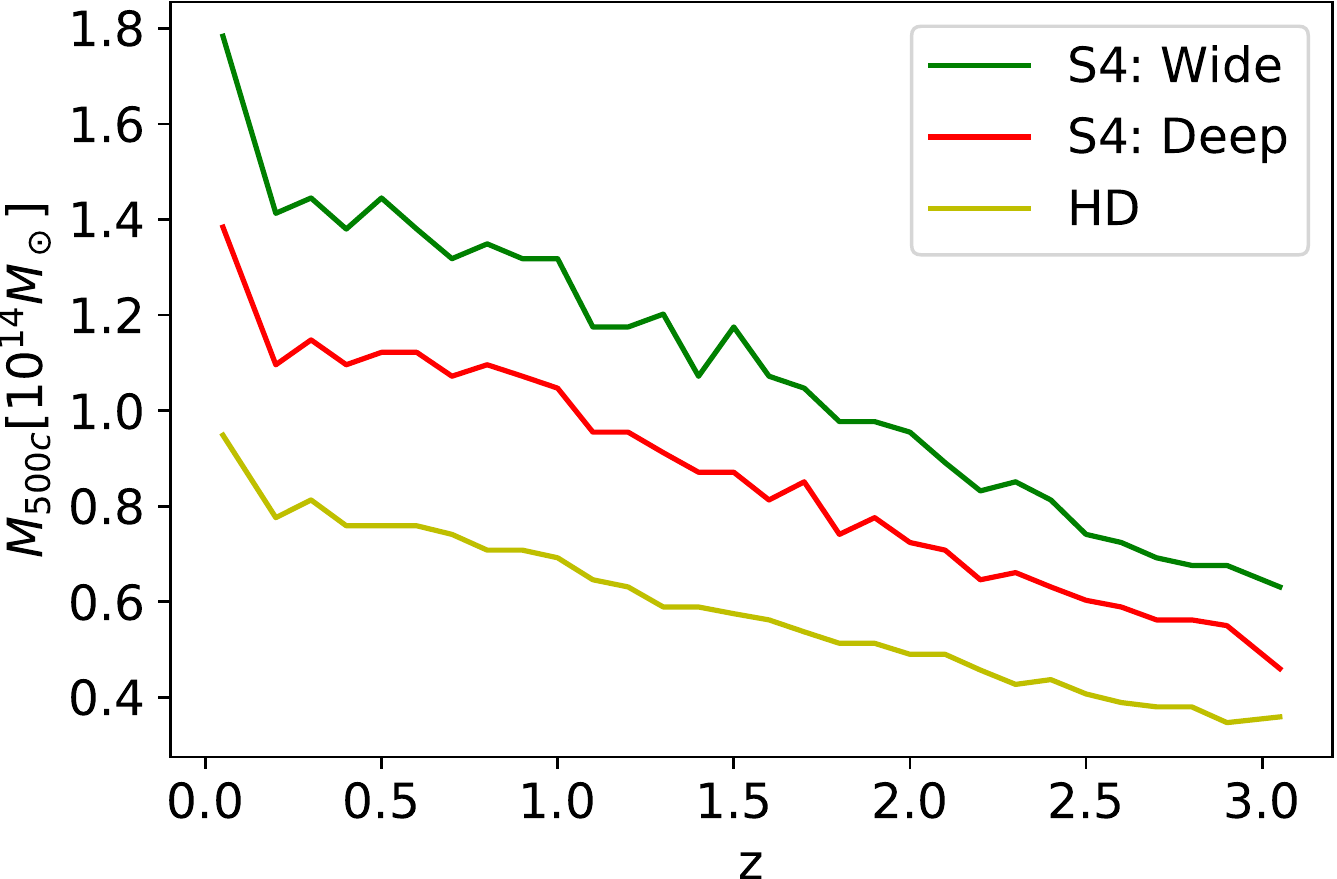}
\end{center}
\setlength\abovecaptionskip{0\baselineskip}
\caption{ The mass detection limit $M_\text{lim}$ as a function of redshift from Ref.~\cite{Raghunathan:2021tdc}, linearly interpolated between their $\Delta z=0.1$ bins.
\label{fig:masslimit}}
\end{figure}

A given experiment will have a selection function in mass and redshift which we approximate here as a simple mass limit as a function of redshift $M_{\rm lim}(z)$.  
For each of the three surveys we forecast, we use the values of $M_{\rm lim}(z)$ calculated in \cite{Raghunathan:2021tdc}. These limits are reproduced in Fig.~\ref{fig:masslimit}. 
The jaggedness of the curves reflects the $\Delta z=0.1$ binning in 
Ref.~\cite{Raghunathan:2021tdc} as does our effective $z_{\rm min} = 0.05$, but we will show in later sections that this effective redshift limit does not affect our results significantly.
The general trend of the  $M_{\rm lim}(z)$ curves---which is the opposite of mass-limit curves from, e.g., X-ray-selected cluster samples---is discussed in Section~3.2.1 of \cite{Raghunathan:2021zfi}.

We model the expected number of total detected clusters for each survey as
\begin{equation}
N_{\rm tot} = 4\pi f_{\rm sky} \int_{z_{\rm min}}^\infty dz \frac {D_A^2(z)}{H(z)} \int_{M_{\rm lim}(z)}^\infty \frac{dM}{M} \frac{dn}{d\ln M} ,
\end{equation}
where $f_{\rm sky}$ is the fraction of sky measured by the experiment. 
We find that for our fiducial cosmology: $N_{\rm tot}=1.04 \times 10^5$ for S4-Wide; $1.10 \times 10^4$  for S4-Deep, and $4.63 \times 10^5$  for CMB-HD.  
Our number of clusters agrees with \cite{Raghunathan:2021tdc} to within $\sim 3\%$ for S4-Wide, $\sim 7\%$ for S4-Deep, and $\sim 10\%$ for CMB-HD.

\subsection{Forecasting Method}

We forecast constraints on $\mu$ from the distorted tSZ spectrum in the direction of massive clusters using a Fisher matrix technique. First, we define the  likelihood per cluster in the catalog. Given the expression for the measured, calibrated tSZ spectrum from Eq.~(\ref{eq:calibratedSZ}), we model the cluster likelihood ${\cal L}$ as
\begin{align}
 -2\ln \mathcal{L}  = & \sum_{i j, \ell m, \ell^\prime m^\prime}
[\Delta T_{i, \ell m} - y_{\ell m} T_0 g(x_i,\mu) ]  \nonumber\\
& {\bf C}^{-1}_{ij,\ell m \ell^\prime m^\prime} [\Delta T_{j, \ell^\prime m^\prime} - y_{\ell^\prime m^\prime} T_0 g(x_j,\mu) ],
\end{align}
where $i$ and $j$ run over frequency bands, $\Delta T_{i, \ell m}$ is the measured, calibrated (spherical harmonic-space) temperature fluctuation in band $i$ in the direction of the cluster, $y_{\ell m}$ is the spherical harmonic-space cluster profile, and we have approximated the sources of noise as Gaussian by characterizing the likelihood with the covariance matrix {\bf C}.
Using Eq.~(\ref{eq:yell}) for the cluster profile and assuming statistical isotropy 
there is no azimuthal dependence in the model or the covariance, and the covariance will be diagonal in $\ell$, in which case we can write
\begin{align}
 -2\ln \mathcal{L}  = & \sum_{i j, \ell}
\frac{2 \ell + 1}{4 \pi}
[\Delta T_{i, \ell} - y({\ell}) T_0 g(x_i,\mu) ]  \nonumber\\
& (\mathbf{C}_\ell)^{-1}_{ij} [\Delta T_{j, \ell} - y({\ell}) T_0 g(x_j,\mu) ].
\end{align}

For the noise covariance matrix, we begin with a baseline of just uncorrelated white noise
and write
\begin{eqnarray}
(\mathbf{C}_\ell)_{ij} \rightarrow (\mathbf{C}_\ell)_{ij}^{\white} &=& \delta_{ij} \frac{C_{\white,i}}{B^2_{\ell,i}},
\label{eq:white}
\end{eqnarray}
where $C_{\white,i}$ is the map noise variance in band $i$, and the Gaussian beam profile is
\begin{equation}
B^2_{\ell,i} \approx \exp \left[- \frac{\ell(\ell+1) }{ {8} \ln 2 }\theta_{{\rm FWHM},i}^{2}  \right].
\label{eq:beam}
\end{equation}
In this case, the likelihood reduces to
\begin{equation}
 -2\ln \mathcal{L}  =  \sum_{i, \ell}
\frac{2 \ell + 1}{4 \pi} \frac{B_{\ell,i}^2}{C_{\white,i}}
[\Delta T_{i, \ell} - y({\ell}) T_0 g(x_i,\mu)]^2.
\end{equation}
More generally we can include other noise terms, indexed by $\rmx$,  
as additional contributions to the covariance matrix 
\begin{equation}
(\mathbf{C}_\ell)_{ij}  = (\mathbf{C}_\ell)_{ij}^{\white} + \sum_\rmx  (\mathbf{C}_\ell)_{ij}^\rmx 
\end{equation}
and in particular for various foreground noise contributions that are fully correlated in frequency space, we take
\begin{equation}
    (\mathbf{C}_{\ell})_{ij}^\rmx = \sqrt{C_\rmx(\ell,\nu_i) C_\rmx(\ell,\nu_j)},
    \label{eq:correlatednoise}
\end{equation}
where $C_\rmx(\ell,\nu_i)$ is the angular power spectrum of component $\rmx$ at frequency $\nu_i$.
We often characterize such contributions using their logarithmic power spectrum
\begin{equation}
D_\rmx(\ell,\nu_i) \equiv \frac{\ell(\ell+1)}{2 \pi} C_\rmx(\ell,\nu_i).
\label{eq:logpower}
\end{equation}

\begin{table*}[bt]
\def\arraystretch{1.5}
\begin{tabular}{|l|c|c|c|}
\hline
\begin{tabular}[c]{@{}l@{}}$\sigma(\mu)$ assuming: 
\end{tabular} & S4-Wide               & S4-Deep               & CMB-HD                \\ \hline
Baseline noise only  & $1.6 \times 10^{-4}$ & $1.4 \times 10^{-4}$ & $2.8 \times 10^{-5}$ \\ \hline
+ 1st order rSZ   & $2.1 \times 10^{-4}$ & $1.9 \times 10^{-4}$ & $3.6 \times 10^{-5}$ \\ \hline
+ CMB $\&$ background kSZ   & $2.5 \times 10^{-4}$ & $2.5 \times 10^{-4}$ & $4.4 \times 10^{-5}$ \\ \hline
+ cluster kSZ   & $2.8 \times 10^{-4}$ & $2.6 \times 10^{-4}$ & $4.6 \times 10^{-5}$ \\ \hline
+ extragalactic foregrounds   & $3.5 \times 10^{-4}$ & $7.0 \times 10^{-4}$ & $1.2 \times 10^{-4}$ \\ \hline
+ galactic foregrounds   & $9.2 \times 10^{-4}$ & $9.1 \times 10^{-4}$ & $1.6 \times 10^{-4}$ \\ \hline
+ atmosphere   & $1.3 \times 10^{-3}$ & $9.9 \times 10^{-4}$ & $1.9 \times 10^{-4}$ \\ \hline
\end{tabular}
\caption{Forecasted constraint on $\mu$ for the
baseline white detector noise of each experimental configuration and its cumulative degradation from additional effects.
\label{tab:challenge}}
\end{table*}

To forecast measurement errors on $\mu$ we employ the Fisher matrix
\begin{equation} 
{{\bf F}}_{\alpha \beta}  = - \left \langle \frac{\partial^{2} \ln \mathcal{L}}{\partial p_{\alpha} \partial p_{\beta}} \right \rangle,
\end{equation}
where in our baseline study we take the parameters as 
$p_\mu \in {y_c,\mu}$ and evaluate the parameter derivatives around a fiducial model with $\mu=0$ and the expected $y_c(M,z)$.  We include $T_e$ as a parameter when considering relativistic corrections in \S\ref{sec:relativisticforecast}.  
In general, the forecasted error on $\mu$ then comes from the $\mu\mu$ element of the matrix inverse of ${\bf F}$,
\begin{eqnarray}
    \sigma_k \left(\mu \right)=\sqrt{({\bf F}^{-1})_{\mu \mu}}\,,
    \label{eq:Fsinglecluster}
\end{eqnarray}
where $k$ indexes the cluster so that the combined result of the independent clusters in the catalogue is given by
\begin{equation}
\sigma^{-2}(\mu) = \sum_k \sigma_k^{-2}(\mu).
\end{equation}

Since the sum over identical clusters involves the same $\sigma_k$, in practice we sum over mass and redshift bins that are  narrow enough so as to provide results that is sufficiently close to the full sum once weighted by the expected number of clusters per bin.

\section{Results}
\label{sec:results}
In this section, we present our main forecasting results. We begin by providing the forecasted constraints on $\mu$ for each of the three experimental configurations in the idealized or ``baseline" case of white detector noise only.
We then introduce real-world complexities that an experiment will have to address, including relativistic corrections to the tSZ effect, CMB background anisotropy, cluster-associated kSZ signal, foreground sources, and atmospheric contamination. We report the degradation of constraints from each of these cumulatively. Since we do not analyze each effect separately, the ordering of the cumulative contributions can matter in the interpretation of which is seemingly the most significant.
We choose this approach to instead emphasize which complexities, in descending order, are fundamental to the measurement and which ones are contaminants to specific experiments.

\subsection{Baseline Noise}

Constraints on $\mu$ for each survey configuration
for the baseline case of white detector noise only 
are shown in the first row of Tab.~\ref{tab:challenge}. 
These represent the most optimistic possible projections from each survey, and the baseline against which we compare the degraded constraints from successive real-world effects in the rest of the table and section.

We notice a few interesting results with regards to our baseline constraints on $\mu$. For the two CMB-S4 surveys in this ideal forecast, $\sigma(\mu)$  is comparable to the bounds from \cobefiras, which constrain $|\mu|<9 \times 10^{-5}$  (95$\%$ CL, \cite{Fixsen:1996nj}). With a CMB-HD-like configuration, we start to see improved constraints on $\mu$ relative to \cobefiras, indicating that, from a raw sensitivity standpoint, this method of constraining $\mu$ has some promise.

\begin{figure*}[htbp]
\setlength\abovecaptionskip{0\baselineskip}
\includegraphics[width=2\columnwidth]{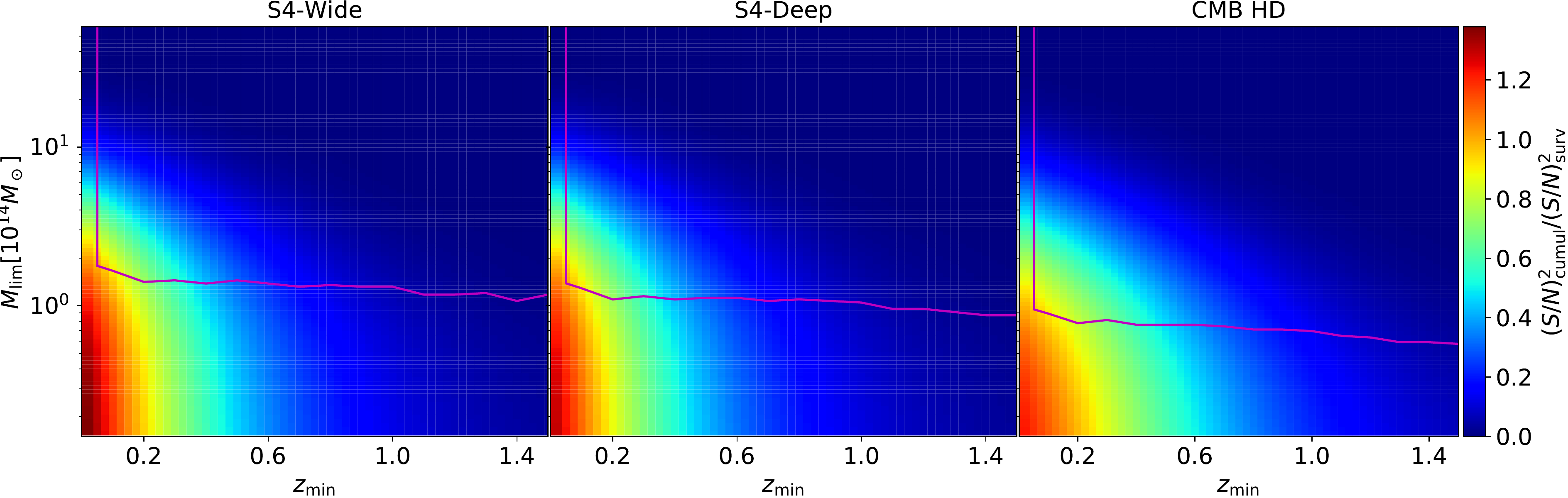}
\caption{The cumulative squared signal to noise, $(S/N)^2_{\rm cumul}$, above a given cluster mass and redshift threshold normalized to the total  $(S/N)^2_{\rm surv}$ for the three surveys from their catalog of clusters with masses and redshifts above the red line.   }
\label{fig:zmSN}
\end{figure*}

In addition, we note that S4-Deep provides a slightly better constraint on $\mu$ than S4-Wide, despite the fact that the constraint comes from an order of magnitude fewer clusters. This can be understood from the fact that, for a fixed set of frequency bands, the per-cluster $\mu \mu$ Fisher matrix element (Eq.~\ref{eq:Fsinglecluster}) 
will scale as the square of the total signal-to-noise $(S/N)$ on the tSZ signal from the cluster. In the ideal white-noise-only case, for the $i$th frequency channel and a cluster of a given mass and redshift, the squared, per-cluster tSZ $S/N$ is given by
\begin{eqnarray}
\label{eqn:snsqr}
\left(\frac{S}{N} \right)_i^2
&=&  \sum_{\ell} \frac{2 \ell + 1}{4 \pi}
\frac{\left[y_{\ell} T_0 g(x_i,0) B_{\ell,i} \right]^2}{C_{\white,i}} \\
\nonumber &=&
\frac{ \left[y_c T_0 g(x_i,0) \right]^2}{C_{\white,i}} \sum_{ \ell} \frac{2 \ell + 1}{ 4 \pi}
{\left[ 2 \pi  \theta _c^2 K_0\left(\ell \theta _c\right) B_{\ell,i} \right]^2} .
\end{eqnarray}
As expected, this scales as $y^2_c/C_{\white,i}$.

The total $\mu$ constraint for a given survey will scale with this quantity summed over all the clusters in the catalog and frequency, $(S/N)_{\rm surv}^2$. For S4-Wide $(S/N)_{\rm surv}^2$ is $1.5 \times 10^9$, for S4-Deep it is $1.9 \times 10^9$, and for CMB-HD it is $2.3 \times 10^{10}$. 
The CMB-S4 Wide survey covers 17 times more sky than the Deep survey, so for any mass and redshift bin above the detection limit of both surveys, the Wide survey will have 17 times more clusters in the catalog. But the square of the ratio of map noise in the main CMB bands in the two surveys---and, by extension the squared $S/N$ per cluster---is over 20, so it is not surprising that the Deep survey attains slightly better $\mu$ constraints.

This line of reasoning ignores the fact that the CMB-S4 Deep survey also has a lower mass limit and a higher cluster number density, which in principle could lead to an even larger difference between the $\mu$ constraints from the Deep and Wide surveys. All of the clusters that will be in the S4-Deep catalog but not the S4-Wide catalog are low-mass systems with $z>z_{\rm min}=0.05$,
but as we shall see next, these clusters do not significantly improve the constraint. 

To understand which clusters are providing most of the constraining power, we calculate the cumulative $(S/N)^2$ above a given mass and redshift and plot that quantity in Figure \ref{fig:zmSN}. 
Specifically we calculate the per-cluster $(S/N)^2$, calculated for frequency channel $i$ using Eq.~(\ref{eqn:snsqr}), sum over frequency channels and clusters above a given mass $M$ and redshift $z$ in the catalogue, and plot this cumulative $(S/N)^2_{\rm cumul}(M,z)$.

Note that Figure~\ref{fig:zmSN} extends the $(S/N)_{\rm cumul}^2$ to below our fiducial values for $M_\text{lim}(z)$ and $z_{\rm min}$ (red lines) so that the ratio with the given survey $(S/N)_{\rm surv}^2$ can exceed unity. 
Nonetheless, in each case half of $(S/N)^2$ at $z_{\rm min}$ comes from cluster
masses well above $M_{\rm lim}(z_{\rm min})$ and at $M_{\rm lim}(z)$ from cluster redshifts below  $z < 0.5$.
This implies that the clusters around 
$M_\text{lim}(z)$ for each survey are not providing much
constraining power on $\mu$ if $z_\mathrm{min}=0.05$. 
It is only for $z<0.05$ and masses substantially below $M_\text{lim}(z_{\rm min})$ that the cumulative $(S/N)^2$ changes noticeably, but even then only by $20-30\%$.\footnote{
At least part of this signal could be recovered by augmenting the internal cluster catalogs with external detections in the optical and X-ray bands. For example, the cluster mass limit for the all-sky survey of the currently operating \textit{eROSITA} mission is $\lesssim 2 \times 10^{13} M_\odot$ at $z < 0.1$ (see, e.g., Figure 5.1.1 in \cite{eROSITA:2012lfj}).}

In this work we produce forecasts for fixed instrument configurations, but it is possible that small modifications to one or more of the configurations could improve the $\mu$ constraints. In particular, it is not obvious from just the total $S/N$ which frequency bands are contributing most to the constraint, and where more bands could potentially help. We note that, when the frequency band allocation is not fixed, the total $\mu$ constraint depends not just on the total $S/N$ but also includes the sensitivity of bands to the $\mu$ distortion. We can write the $\mu \mu$ Fisher matrix element as
\begin{equation}
{\bf F_{\mu\mu}} = \sum_i \left(\frac{S}{N} \right)_{i}^2 \left( \frac{\partial\ln g(x_i,\mu)}{\partial \mu} \right)^2
\end{equation}
where recall $\partial\ln g/\partial \mu$ is shown in Fig. \ref{fig:muy}.
Of course, the final constraint on $\mu$ depends on the other contributions to the signal that must be marginalized over.

In the simple case where only $y_c$ is marginalized over, we can build intuition 
for which frequencies contribute most to the $\mu$ constraint
by considering the scenario with only two channels, in which case the squared uncertainty on $\mu$ (or, equivalently, the $\mu \mu$ element of the inverse Fisher matrix) is given analytically by
\begin{equation}
\sigma^2(\mu) = \frac{(S/N)^2_1+ (S/N)^2_2}
{(S/N)^2_1 (S/N)^2_2}
\left[ 
\frac{\partial}{\partial\mu} \ln\frac{g(x_1,\mu)}{g(x_2,\mu)}\right]^{-2}.
\label{eq:channels}
\end{equation}
From Eq.~(\ref{eq:channels}), it is clear that both sensitivity and frequency lever arm are important for constraining $\mu$, as $\sigma^2(\mu)$ blows up when the $S/N$ in either of the two bands gets too small or when $g(x,\mu)$ is similar enough between the bands that the spectral signature becomes indistinguishable from that of $y_c$. 
For estimation purposes, we find that the expression 
\begin{equation}
\left(\frac{S}{N} \right)^2_{i} \approx \frac{(y_c T_0 )^2}{C_{\white,i}}
\frac{\pi \theta_c^2 g^2(x_i,0)}{1+{(\theta_{{\rm FWHM},i}/4\theta_c)}^{1.6}}
\end{equation}
approximates Eq.~(\ref{eqn:snsqr}) to within a few percent for all clusters and instrument configurations discussed here.
To further illuminate scaling results we can also roughly scale  $\sqrt{C_\white}$ and $\theta_\text{FWHM}$ with frequency from 150GHz to mimic CMB-S4 Wide survey specifications:
\begin{align}
\label{eq:cwhite}
 \frac{\sqrt{C_{\white}}(\nu)}{2 \; \mu\textrm{K-arcmin}} ={}& 1 + 11.5\left(\frac{\nu}{150\; \textrm{GHz}}-1\right)^2, \\
\label{eq:beamscale}
\frac{\theta_\mathrm{FWHM}(\nu)}{1.5 \; \text{arcmin}} ={}& \left(\frac{\nu}{150 \; \mathrm{GHz}}\right)^{-1}.
\end{align}

\begin{figure}[t]
\begin{center}
\includegraphics[width=1\columnwidth]{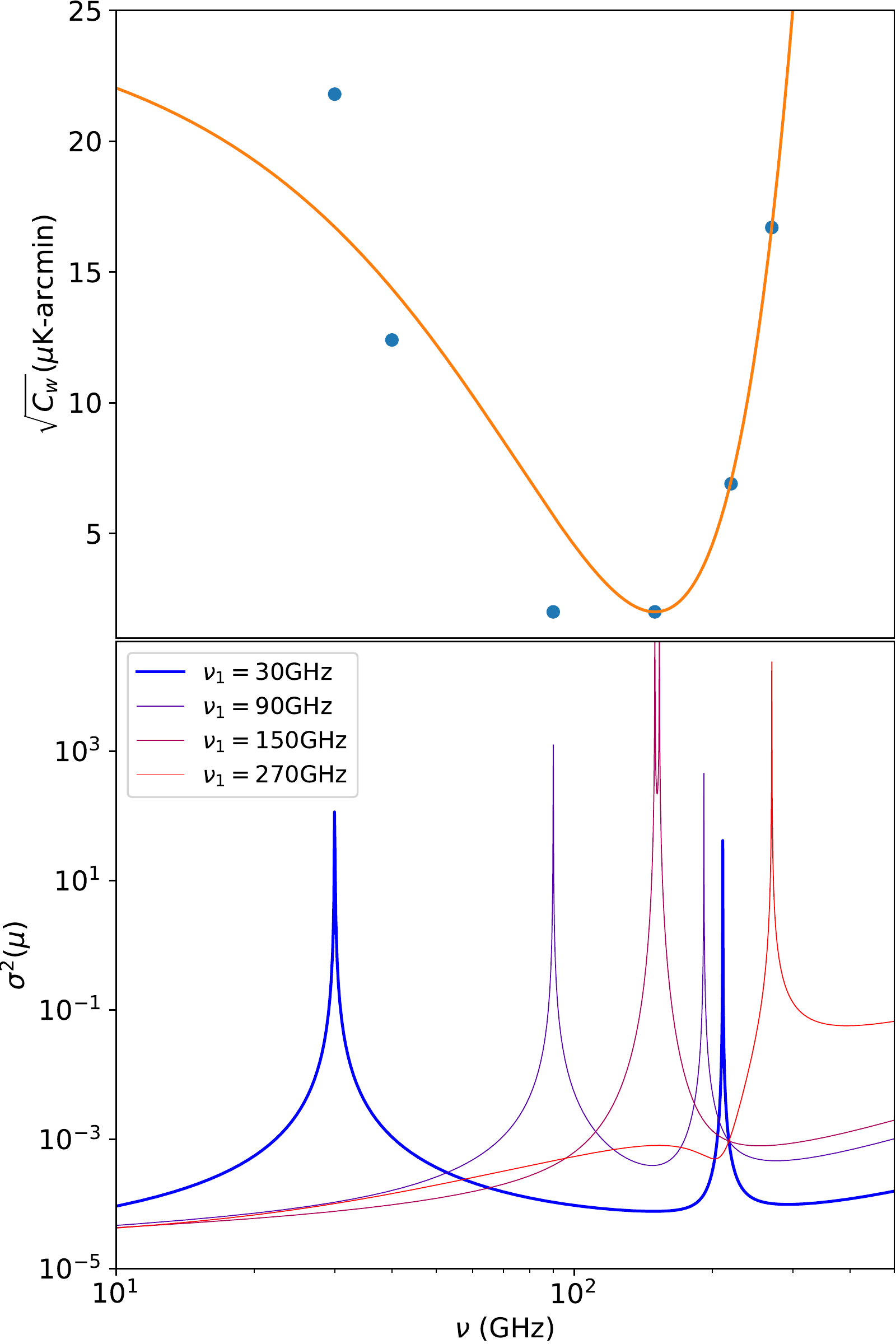}
\end{center}
\setlength\abovecaptionskip{0\baselineskip}
\caption{Top: Frequency scaling relation for $\sqrt{C_{\white}}$. Blue points correspond to the S4-Wide survey's specifications. Bottom: $\sigma^2(\mu)$ for different frequency pairs for a single cluster with $y_c = 10^{-4}$ and $\theta_c = 1$ arcminute. We see that the constraints on $\mu$ improve when the channels are separated from each other, rather than being closely spaced. In general, we see that as long as there is a separation, lower frequency channels provide more sensitivity to $\mu$ in this baseline case of white detector noise only,  in accordance with Fig.~\ref{fig:muy}.
\label{fig:weights}}
\end{figure}

We plot Eq.~(\ref{eq:cwhite}) in Fig.~\ref{fig:weights} (upper panel, curve) and compare it against the actual CMB-S4 Wide channel noise (points). 
Using this noise curve and Eq.~(\ref{eq:beamscale}),  in Fig.~\ref{fig:weights} (bottom panel), we show  $\sigma^2(\mu)$ 
for a cluster with $y_c = 1 \times 10^{-4}$ and $\theta_c = 1$ arcminute in this two-channel case,  as a function of the frequency of the second channel $\nu$ with the first fixed at either $\nu_1=$\,30, 90, or 150\;GHz.

Notice that $\sigma^2(\mu)$ diverges whenever the two frequencies have the same value of $\partial \ln(\Delta T) / \partial \mu$ (see Fig.~\ref{fig:muy}), causing $\mu$ to become degenerate with $y_c$ in the fit.  This  occurs by definition when the two frequencies are coincident, and, for a lower frequency $\nu_1$ below the null, it occurs again for a specific upper frequency $\nu_2$.  
In the limit where the lower frequency goes to zero and the $\mu$ response diverges, this second degeneracy between $\mu$ and $y_c$ occurs when the upper frequency approaches the null.  For a lower frequency around 150\;GHz, the degeneracy disappears since the response in Fig.~\ref{fig:muy} is near the local minimum where it is single-valued in frequency.  The degeneracy is ``accidental" in the sense that it only exists for pairs of frequency channels and is resolved  once there are three or more channels.  As we shall see next, the more complexity we add onto this baseline case the more multiple frequency channels are required to distinguish the $\mu$ signal.

\subsection{Relativistic corrections}
\label{sec:relativisticforecast}

So far when forecasting $\mu$-distortions from the distorted tSZ spectrum, we have used the non-relativistic limit of the tSZ frequency spectrum. In reality, the hottest clusters, from which most of our constraining power is derived, are going to have non-negligible relativistic corrections, sometimes called the relativistic Sunyaev-Zeldovich (rSZ) effect,  especially compared to the small level of distortion that $\mu$ introduces. We show in Appendix \ref{sec:relativistic} that the $\mu$-distorted rSZ spectrum modifies Eq.~(\ref{eq:calibratedSZ}) for the anisotropy calibrated temperature fluctuation to 
\begin{equation}
    \Delta T= y T_0   g(x,\mu,\theta_e ),
\end{equation}
where $\theta_e=k_B T_e/m_e c^2$.   This generalizes from the form $g(x,\mu,0)= g(x,\mu)$ given in Eq.~(\ref{eq:g}).

\begin{figure}[t!]
\begin{center}
\includegraphics[width=1\columnwidth]{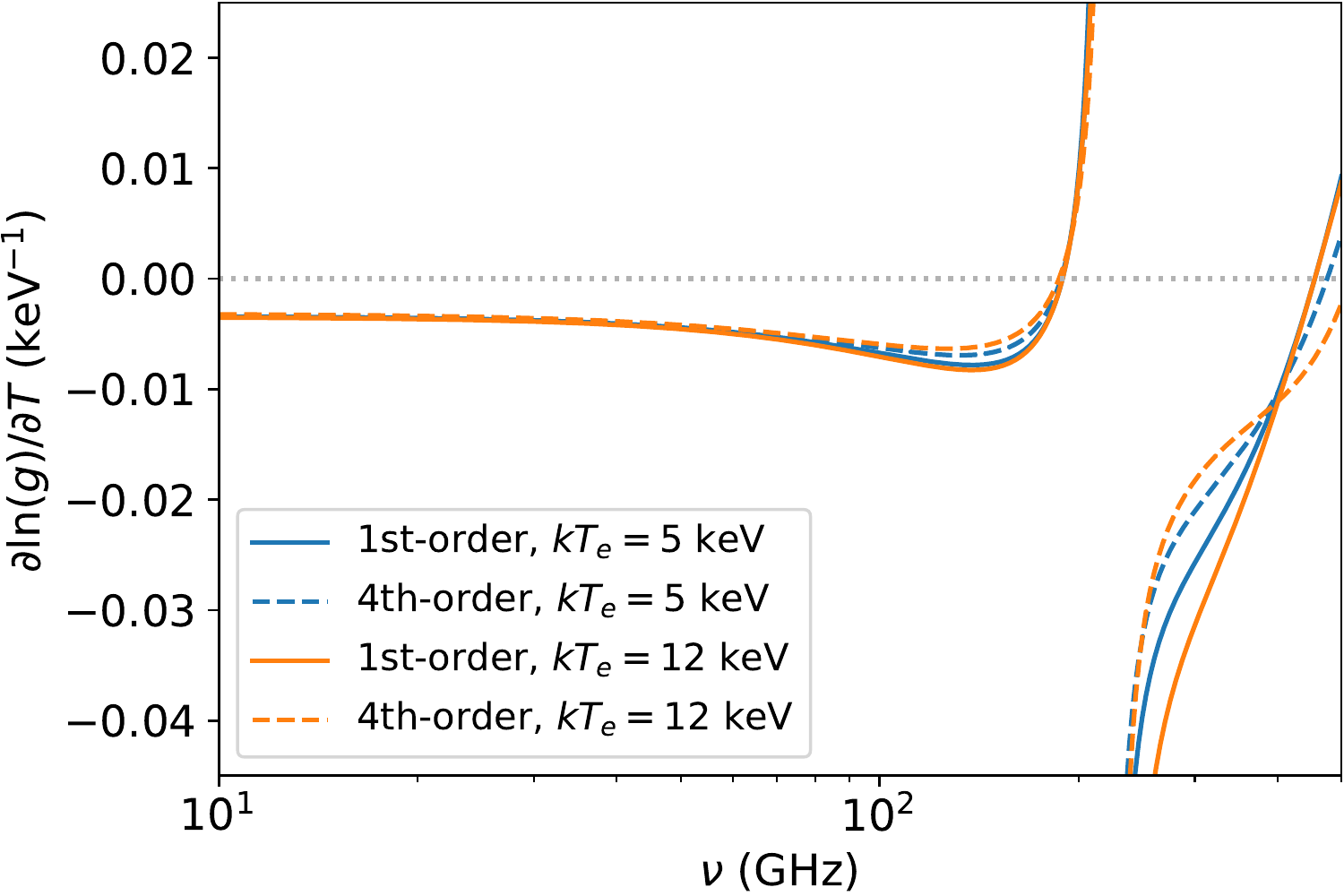}
\end{center}
\caption{The fractional response of the tSZ spectrum, with 1st-order and 4th-order relativistic corrections, to the temperature of the cluster. Here we take $\mu=0$ for a cluster with $k_B T_e = 5$ keV and another with $k_B T_e = 12$ keV. 
\label{fig:rel}}
\end{figure}

We then marginalize over $T_e$ per cluster bin, around the central values given by Eq.~(\ref{eq:temp}), along with $y_c$ using the first order in $\theta_e$
expression for $g$ from Eq.~(\ref{eq:ggen}).  In Fig.~\ref{fig:rel}, we show the corresponding fractional change in the $y$ distortion as a function of frequency for a range of cluster temperatures for comparison with Fig.~\ref{fig:muy} for the $\mu$ distortion.  Marginalizing over $T_e$ per cluster bin degrades our constraint on $\mu$ by $\sim 30\%$ (see Tab.~\ref{tab:challenge}).
While this is not a large effect by itself, marginalizing over $T_e$ has the effect of using up another linear combination of frequency bands to help break the degeneracy with $\mu$, as was the case with $y_c$.

As noted by Ref.~\cite{Itoh:1997ks}, the convergence of relativistic corrections as a Taylor expansion in powers of $\theta_e$ is slow at frequencies around the null and above.  In Fig.~\ref{fig:rel}, we also show the spectral shape of the relativistic correction at 4th order. The small change in the shape associated with the central frequencies of the surveys, which are below the null,
implies a correspondingly small change in the $\mu$ constraints.  We find that going to 4th order makes a $5 \%$ change for S4-Wide, a $6\%$ change for S4-Deep, and a $3\%$ change for CMB-HD.

\subsection{CMB anisotropies}
\label{sec:cmb} 
Another source of variance in the measurement of the tSZ effect from clusters is primary CMB anisotropy. As can be inferred from Fig.~\ref{fig:zmSN}, the constraint on $\mu$ is dominated by high-mass, low-redshift clusters. These clusters are sufficiently extended on the sky that primary anisotropy is a potential concern.

There are also secondary CMB anisotropies arising from the kinetic Sunyaev-Zeldovich (kSZ) effect due to scattering off of gas through its bulk rather than thermal motion after recombination. The kSZ effect is intrisincally a Doppler shift and has the same spectrum as primary anisotropies (see \S \ref{sec:calibration}).
In this section, we only consider the background kSZ signal rather than the contribution specific to the clusters in our catalog. The isotropic kSZ signal is subdominant relative to the primary CMB anisotropies until $\ell \gtrsim 4000$.

Since these sources are statistically isotropic, we model their effects on our constraints by including the CMB temperature power spectrum and kSZ power spectrum in our noise covariance matrix. As described in \S\ref{sec:mu}, the spectrum for CMB anisotropies will look like the derivative of a blackbody given that our experiments will calibrate off the anisotropies.  Such sources  have a constant temperature across frequencies by definition. This means that the temperature 
power spectra will act as frequency independent, fully-correlated noise across frequency channels in Eqs.~(\ref{eq:correlatednoise}), (\ref{eq:logpower}).
We use CAMB to generate the primary CMB logarithmic power spectrum $D_{TT}({\ell})$ for the fiducial cosmology. For the kSZ power spectrum, we consider a scale invariant spectrum with a constant $D_\text{kSZ}({\ell})$. We use the amplitude measured from the South Pole Telescope (SPT) at $\ell = 3000$ \cite{George:2014oba},  $D_\text{kSZ}({3000}) = 2.9 \ \mu \text{K}^2$.

As shown in Tab.~\ref{tab:challenge}, $\sigma(\mu)$ slightly degrades when including the CMB and kSZ background, but not by a large amount. For both CMB-S4 surveys, the individual contribution of the CMB and kSZ effect are similar to one another. The smaller beams and wider $\ell$ coverage of CMB-HD make the kSZ effect relatively more important than the primary CMB.

\subsection{Cluster Kinetic Sunyaev-Zeldovich Effect}

In the previous section, we only considered the kSZ signal corresponding to a statistically isotropic background.
Since each cluster also has a specific kSZ profile associated with its gas profile and peculiar motion, we cannot treat it as statistically isotropic noise as we can with foregrounds that are not associated with the cluster. 

We could model the kSZ signature from each cluster 
as part of the cluster signal and marginalize over the peculiar velocity of the cluster as we did for $y_c$ and $T_e$.   On the other hand, 
the spectrum of the kSZ is perfectly known, and we can effectively marginalize over a signal with a known spectrum by adding a component with that spectrum and artificially high amplitude to the covariance matrix \cite{Bond:1998zw}. Since kSZ has the same spectrum as the primary CMB anisotropy, which already is in the covariance matrix with an amplitude much larger than the instrumental noise, we expect this procedure to have minimal impact on our $\mu$ constraints.
In practice, we marginalize over any signal with the spectrum of primary CMB anisotropy or kSZ by multiplying the 
$D_{TT}+D_{\textrm{kSZ}}$ spectra by a sufficiently large constant that the resulting $\sigma(\mu)$ saturates to its asymptotic value.  Any part of the $\mu$-signal that comes from the combination of frequency channels with a blackbody spectrum will have effectively infinite noise and not contribute to the constraint.

The fourth row of Table \ref{tab:challenge} shows $\sigma(\mu)$ when we implement this procedure. We see that our constraints hardly change for all three CMB surveys. This suggests that, as expected, any contribution to the $\mu$ constraint from the combination of frequency band information corresponding to the spectrum of primary CMB anisotropy is already made negligible by the inclusion of the fiducial $D_{TT}+D_{\textrm{kSZ}}$ in the covariance.

\subsection{Extragalactic foregrounds}

We treat three independent types of extragalactic foregrounds $\rmx \in {\clustered, \poisson, \radio}$: ``$\clustered$'', the clustered cosmic infrared background (CIB); ``$\poisson$'', the spatially unclustered or ``Poisson'' component of the CIB; and  ``$\radio$'', the radio sources or synchrotron-emitting active galactic nuclei, the clustering of which is assumed to be negligible. 
For extragalactic foregrounds and, in the next section, galactic foregrounds, we parameterize the logarithmic power spectrum of foreground $\rmx$ at multipole $\ell$ and frequency $\nu$ as
\begin{equation}
D_\rmx(\ell,\nu) \propto \left[ \frac{ f_\rmx(\nu)}{( \partial B/\partial T)_\nu}\right]^2 \ell^{\beta_\rmx},
\end{equation}
and provide the normalization  $D_\rmx(\ell_\rmx,\nu_\rmx)$ at a fiducial multipole value $\ell_\rmx$ and frequency $\nu_\rmx$.  Here $\beta_\rmx$ is the index of the assumed power-law multipole dependence, $f_\rmx(\nu)$ encodes the frequency dependence in specific intensity units, 
and $(\partial B/\partial T)_\nu$ converts specific intensity to CMB temperature units at the relevant frequency $\nu$ assuming a blackbody CMB spectrum.\footnote{Note that we neglect
the $\mu$-dependent anisotropy calibration $C$-factors from Eq.~(\ref{eq:miscalibration}) here, since we are considering the foregrounds as noise rather than signal.  In principle measuring anisotropy-calibrated foregrounds  with  known absolute spectra could themselves be used to measure $\mu$.}
We assume 100\% correlation between observing bands for all individual extragalactic and galactic foreground components, such that we only need to specify the behavior at a single frequency, with the cross-frequency components of the covariance given by Eq.~(\ref{eq:correlatednoise}).

For all three extragalactic foregrounds considered here, we assume a power-law frequency dependence 
\begin{equation}
f_\rmx(\nu) = \nu^{\alpha_\rmx}.
\label{eqn:plaw}
\end{equation}
Following \cite{George:2014oba}, for the clustered CIB we adopt 
$D_\clustered(3000,150\,\GHz) = 3.5 \, \mu \mathrm{K}^{2}$ and $\alpha_\clustered$ = 4.3,
while for the Poisson component of the CIB we adopt $D_\poisson(3000,150\,\GHz) = 9.2 \, \mu \mathrm{K}^{2}$ and $\alpha_\poisson = 3.3$, and for
the radio Poisson component, we adopt $D_\radio(3000,150\,\GHz) = 2.0 \times 10^{-2} \, \mu \mathrm{K}^{2}$ and $\alpha_\radio = -0.70$.
For the $\ell$ dependence, we adopt $\beta_\clustered = 0.80$ (again following \cite{George:2014oba}), while for the Poisson terms $\beta_\poisson=\beta_\radio=2$ by definition.

We note that the assumed radio amplitude we adopt is significantly lower than the best-fit radio source power quoted in \cite{SPT:2020psp} (the follow-up paper to \cite{George:2014oba}, which did not constrain the radio amplitude), owing to the assumption of a much lower source cut threshold in the experiments treated here. The $5 \sigma$ point source threshold in the 150~GHz channel of the CMB-S4 Deep survey should be roughly $50 \times$ lower than that used in \cite{SPT:2020psp}, and the slope of the number counts of radio sources is such that the Poisson power should scale roughly linearly with flux cut. As such, we adopt a radio amplitude value $50 \times$ smaller than the $1.0 \; \mu \mathrm{K}^{2}$ value from \cite{SPT:2020psp}. This number will be slightly optimistic for CMB-S4 Wide and slightly pessimistic for CMB-HD.

We see in Tab.~\ref{tab:challenge} that extragalactic foregrounds have a larger impact on  our constraints of $\mu$ than the effects of the previous sections. Moreover CMB-HD no longer provides improvements on $\mu$ compared to the constraints from \cobefiras. More specifically, in all three of our surveys, radio point sources are the dominant extragalactic foreground contaminating our constraints on $\mu$. Both the clustered and Poisson contributions of the CIB have a negligible effect on $\sigma(\mu)$. Thus, efforts to reduce the effects of extragalactic foregrounds should prioritize mitigating the effects of radio sources. One clear path forward is to exploit the available lower-frequency data, both in the CMB surveys themselves and in planned contemporaneous radio surveys such as the Square Kilometer Array (SKA, \cite{Weltman:2018zrl}), the source detection threshold of which will be such that masking of SKA-detected sources in CMB-S4 or CMB-HD data will be limited by the number of independent pixels or resolution elements in the map. Using current source models (e.g., \cite{DeZotti:2004mn}), the source density at flux cut levels a factor of several lower than those assumed here still only reaches hundreds per square degree, still feasible for masking in CMB-HD data.

Note that we are implicitly treating extragalactic foregrounds as a statistically isotropic background to the cluster signal. In fact, galaxy clusters will likely be overdensities of ``foreground'' contamination as well as the desired tSZ signal. A potential method to account for this would be to parameterize the covariance matrix with amplitude parameters for each extragalactic component and marginalize over these parameters per cluster. With sufficiently informative priors from observations in other surveys and at other wavelengths, this could be achieved with a minimal degradation of the eventual $\mu$ constraint.

\subsection{Galactic foregrounds}

The sources of galactic contamination that have traditionally been considered most important at CMB observing frequencies are thermal dust emission and synchrotron emission, but we also include a component of ``anomalous microwave emission" (AME) because of its importance at low frequencies (e.g., \cite{Dickinson:2018rle}).

Because galactic foregrounds are not statistically isotropic, we adopt separate sets of values for foreground amplitudes for the $f_\text{sky}=0.03$ S4-Deep survey and the $f_\text{sky} = 0.5$ region targeted by the S4-Wide survey (which we also adopt for the CMB-HD survey).\footnote{Because of the statistically anisotropic nature of the galactic foregrounds, in a real data set, the covariance for clusters in different parts of the sky would be potentially quite different, and using the mean covariance across the sky for all clusters is not strictly correct.}

Interstellar dust heated by starlight emits as a quasi-thermal modified blackbody. We follow Ref.~\cite{CMB-S4:2020lpa} and parameterize the frequency behavior 
of thermal dust emission as
\begin{equation}
    f_\dust(\nu) = \nu^{\alpha_\dust} B_\nu(T_\dust),
\end{equation}
where $T_\dust = 19.6 \mathrm{K}$ is the dust temperature, and $\alpha_\dust = 1.6$. Also following that work, we set $\beta_\dust=-0.4$.
Following Ref.~\cite{Dibert:2022gep}, we adopt $D_\dust(80,145\,\GHz)=3.3 \, \mu \mathrm{K}^{2}$ for S4-Deep and $1.2 \times 10^3 \, \mu \mathrm{K}^{2}$ for S4-Wide and CMB-HD.\footnote{This is technically for $f_\mathrm{sky}=0.58$, but if we recalculate for $f_\mathrm{sky} = 0.50$, the value only decreases by $\sim 30\%$.} This very large increase from Deep to Wide is attributed at least partly to the requirement adopted in the CMB-S4 Wide survey to restrict observing elevation to $\ge 40^\circ$. If we impose no elevation restriction and instead choose the 50\% of the sky at highest galactic latitude (using the publicly available {\tt PySM} simulations \cite{Thorne:2016ifb} as in \cite{Dibert:2022gep}), we find $D_\dust(80,145\,\GHz)=63 \, \mu \mathrm{K}^{2}$. This would similarly reduce the impact of AME on the wide surveys (see below for details).

Again following Ref.~\cite{CMB-S4:2020lpa}, we parameterize synchrotron as a pure power law in frequency (as in Eq.~\ref{eqn:plaw}) and adopt $\alpha_\sync = -1.1$ and $\beta_\sync = -0.4$. Likewise following Ref.~\cite{Dibert:2022gep}, we adopt 
$D_\sync(80,93\,\GHz)=5.0 \times 10^{-3} \, \mu \mathrm{K}^{2}$ for S4-Deep and $5.5 \times 10^{-2} \, \mu \mathrm{K}^{2}$ for S4-Wide and CMB-HD. We note that the synchrotron amplitude does not vary as strongly across the sky in {\tt PySM} as the dust amplitude: The ratio of power in the Wide and Deep areas is only $\sim 10$ for synchrotron, compared to over 300 for dust. Similarly, if we use $|b| > 30^\circ$ instead of the official CMB-S4 Wide region, we find that the synchrotron amplitude decreases by less than a factor of two (compared to $\sim 20$ for dust).

Because of the potential importance of low-frequency information in our $\mu$ constraint, we also consider the impact of AME.
We investigate the behavior of AME in the CMB-S4 3\% sky region using {\tt PySM}. We find that the AME SED has a double-peaked shape, which we parameterize as
\begin{equation}
f_\ame^2(\nu) =
{e^{-\left[\ln(\nu/\nu_1)\right]^2/2 \sigma_1^2} + A e^{-\left[\ln(\nu/\nu_2)\right]^2/2 \sigma_2^2}} ,
\end{equation}
with $\nu_1 = 10\,\mathrm{GHz}$, $\sigma_1 = 0.43\,\mathrm{GHz}$, $\nu_2 = 22\,\mathrm{GHz}$, $\sigma_2 = 0.35\,\mathrm{GHz}$, and $A = 6.5 \times 10^{-3}$. 
We assume $\beta_\ame = -0.4$ (as would be expected if AME were from spinning dust grains and traced the thermal dust emission). From {\tt PySM}, we estimate $D_\ame(80,10\,\GHz)=1.0 \times 10^4 \, \mu \mathrm{K}^{2}$  for the S4-Deep survey and, assuming the same scaling between deep and wide found for the thermal dust, $D_\ame(80,10\,\GHz)=3.6 \times 10^6 \, \mu \mathrm{K}^{2}$ for S4-Wide and CMB-HD.

The inclusion of galactic foregrounds has a larger impact on S4-Wide compared to S4-Deep and CMB-HD, which are more impacted by extragalactic than galactic foregrounds due to their $\mu$ constraint being weighted toward higher multipoles (see Tab.~\ref{tab:challenge}). More specifically, S4-Wide and S4-Deep now have comparable constraints on $\mu$, despite vastly different galactic foreground amplitudes. Most notably, AME is responsible for most of the degradation in $\sigma(\mu)$ for all three surveys. Synchrotron provides some contribution, while being sub-dominant to AME, and dust has a negligible effect on $\sigma(\mu)$. As discussed above, relaxing restrictions on observing elevation in the wide surveys can help mitigate the impact of AME. But this result also motivates a more careful investigation into the spectral and spatial behavior of AME, beyond the simple ansatz made in this work.

\subsection{Atmosphere}

We saw in our $S/N$ contour plot that most of the signal is from high-mass, low redshift clusters, and they can subtend a large angle in the sky. 
In addition to being potentially confused with primary CMB fluctuations (see \S~\ref{sec:cmb}), signals from objects this large on the sky are also impacted (in ground-based measurements) by emission from poorly mixed water vapor in the atmosphere. The amplitude of water-vapor fluctuations in the atmosphere is higher at large spatial scales than small spatial scales, and the emission thus behaves as ``red noise'' in CMB maps, often modeled as a power law in $\ell$.
The total detector + atmosphere noise power in frequency band $i$ can then be parameterized with three numbers, namely the white noise level $C_\white$, the multipole value at which the detector and atmosphere noise levels are equal $\ell_{\rm knee}$, and the power-law index of the atmosphere noise $\alpha_{\mathrm{atmo}}$:

\begin{equation}
C_{\white,i} \rightarrow C_{\white,i}
     \left [ 1+ \left ( \frac{\ell_{\mathrm{knee},i}}{\ell} \right )^{\alpha_{\mathrm{atmo},i}} \right ] .
\label{eqn:uatmo}
\end{equation}
Our values of $\ell_{\mathrm{knee}}$ and $\alpha_{\mathrm{atmo}}$ for the three surveys are taken from \cite{Raghunathan:2021tdc} and given in Tab~\ref{tab:atmosphere}.

\begin{table}[]
\def\arraystretch{1.5}
\setlength{\tabcolsep}{4pt}
\begin{tabular}{|l|c|cccccc|}
\hline
\multicolumn{2}{|c|}{Channels (GHz)}                           & 30   & 40    & 90    & 150   & 220 & 270        \\\hline \cline{1-8} 
\multicolumn{1}{|c|}{Survey}           &          $f_\text{sky}$      & \multicolumn{6}{c|}{$\ell_{\mathrm{knee}}$ \& $\alpha_{\mathrm{atmo}}$ } \\ \hline
S4-Wide                          & 50\%           & 400  & 400  & 1900  & 3900  & 6700 & 6800   
\\ &
 & 3.5    & 3.5        & 3.5       & 3.5       & 3.5        & 3.5       \\ \hline
S4-Deep                          & 3\%            & 400  & 400 & 1200   & 1900   & 2100 & 2100        
\\ &
& 4.2  & 4.2 & 4.2 & 4.1 & 4.1   & 3.9      \\ \hline
CMB-HD                           & 50\%           & 400  & 400  & 1900  & 3900  & 6700 & 6800   
\\ &
 & 3.5    & 3.5        & 3.5       & 3.5       & 3.5        & 3.5       \\ \hline
\end{tabular}
\caption{Atmosphere parameters for the CMB-S4 Wide and Deep surveys and a CMB-HD-like survey, taken from \cite{Raghunathan:2021tdc}.}
\label{tab:atmosphere}
\end{table}

While Eq.~(\ref{eqn:uatmo}) describes atmospheric emission as uncorrelated between frequency bands, physical intuition and empirical evidence (e.g., \cite{Holzapfel:1997pa}) argue that it should in fact be strongly correlated between bands, at least for instruments in which the beam patterns for detectors at different frequencies overlap in the atmosphere.
The effects of atmosphere could in principle be reduced using the correlation between frequency bands to project out much of the atmospheric contamination.

\subsection{Order of Operations}

\begin{table}[b]
\def\arraystretch{1.5}
\begin{tabular}{|l|c|c|c|}
\hline
\begin{tabular}[c]{@{}l@{}}Excluded 
\end{tabular} & S4-Wide               & S4-Deep               & CMB-HD                \\ \hline
1st order rSZ  & $\times 0.60$ & $\times 0.58$ & $\times 0.68$ \\ \hline
CMB $\&$ all kSZ   & $\times 0.90$ & $\times 0.94$ & $\times 0.97$ \\ \hline
cluster kSZ   & $\times 0.97$ & $\times 0.99$ & $\times 0.98$ \\ \hline
extragal. fore.   & $\times 0.99$ & $\times  0.79$ & $ \times 0.92$ \\ \hline
gal. fore.   & $\times 0.41$ & $\times 0.83$ & $\times 0.71$ \\ \hline
\end{tabular}
\caption{ Forecasted fractional improvement to $\sigma(\mu)$ (see last line of Tab.~\ref{tab:challenge} for the values for each experimental configuration) when excluding certain individual effects. 
\label{tab:order}}
\end{table}

 Our chosen ordering of cumulative effects may give the impression that certain effects are negligible because they are when implemented early in the ordering. However, these effects could prove significant when implemented last, after the survey's constraining power is used to fix other effects. To help gauge the impact each effect has on the end result, we calculate $\sigma(\mu)$ when excluding individual effects from the end result. These results are shown in Tab.~\ref{tab:order}.

$\sigma(\mu)$ when excluding relativistic contributions, compared to including 1st-order corrections, improves by $\sim 40 \%$ for both CMB-S4 surveys, whereas the results improve by $\sim 30\%$ for CMB-HD. 
If we include the effect up to 4th order, the results are almost identical to 1st-order results.

The CMB and the kSZ effect, both the isotropic and cluster component, have a negligible impact on $\sigma(\mu)$ when excluded at the end.
Interestingly, extragalactic foregrounds also have a negligible impact on $\mu$ for S4-Wide and CMB-HD. S4-Deep's constraint improves when we exclude extragalactic foregrounds, but only by about $25\%$. 

The exclusion of galactic foregrounds improves S4-Wide and CMB-HD's constraints on $\mu$, but only modestly improves S4-Deep's constraints. This is likely because contamination from galactic foregrounds is much worse for S4-Wide and CMB-HD, which include observations near the galactic place. 

Our results suggest that the largest way to improve constraints on $\mu$ for all surveys is to address relativistic corrections. For S4-Wide and CMB-HD, galactic foregrounds are a major challenge to improving constraints. For S4-Deep, galactic and extragalactic foregrounds present similar levels of degradation. Generally, addressing these challenges requires additional frequency channels in order to help isolate the $\mu$ signal. Finally, we note that priors on $T_e$ can in principle be obtained from external data such as X-ray observations.

\subsection{Interfrequency Calibration Requirement}

While we can effectively perfectly account for the mis-calibration induced from assuming the background photon distribution is a blackbody when it is in fact a Bose-Einstein distribution, in a real instrument there will also be mis-calibration from the fact that the observation of the calibration source is not noise-free. If we parameterize this calibration error as 
\begin{equation}
C_\mathrm{meas} = C_\mathrm{true} (1 + \dcal) \equiv 1 + \dcal,
\end{equation}
then in a real experiment, the measured, (mis-)calibrated signal from a single cluster will be 
\begin{equation}
\Delta T(x, y_c, \mu, \dcal) = y T_0 g(x,\mu)  (1 + \dcal).
\end{equation}
The requirement for interfrequency calibration is most obvious in the Rayleigh-Jeans limit (and in the limit $\mu \ll x$), in which $g(x,\mu) = -2 (1 + \mu/x)$. The basic information used to constrain $\mu$ is the ratio of the cluster signal in two bands. If we assume perfect calibration in one band and a mis-calibration in the other, we find
\begin{eqnarray}
R (x_1, x_2, y_c, \mu, \dcal) &=& \frac{\Delta T(x_2, y_c, \mu, \dcal)}{\Delta T(x_1, y_c, \mu)} \\
\nonumber &=& \frac{-2 y_c T_0 (1 + \mu / x_2) (1 + \dcal)}{-2 y_c T_0 (1 + \mu / x_1)} \\
\nonumber &=& \frac{1 + \dcal + \mu / x_2 + \dcal \mu / x_2}{1 + \mu / x_1} \\
\nonumber &\simeq& \left( 1 + \dcal + \frac{\mu}{x_2} \right)
\left( 1 - \frac{\mu}{x_1} \right) \\
\nonumber &\simeq& 1 + \mu \left( \frac{1}{x_2} - \frac{1}{x_1} \right) + \dcal.
\end{eqnarray}

It is clear from this formulation that to constrain $\mu$ to some level $\sigma(\mu)$, we need calibration uncertainty smaller than $\sigma(\mu) \left | (x_2^{-1} - x_1^{-1}) \right |$. For the experimental configurations considered in this work, that means we need calibration better than $10^{-4} - 10^{-5}$ in the bands around the peak of the CMB blackbody. A full-sky experiment with noise levels of 1 $\mu$K-arcmin has  $S/N$ per band on the primary CMB temperature anisotropy approaching $10^{6}$, so in principle this level of interfrequency calibration is achievable. 

Additionally, this level of calibration precision must be maintained over the full survey area. For surveys that cover a large fraction of the sky, different parts of the survey are in general surveyed at widely separated times and possibly under different atmospheric conditions. This places an effective requirement on calibration stability; alternatively, different parts of the survey can be calibrated independently, in which case the $S/N$ requirement on the CMB is per independently calibrated patch.

Finally, we note that this calibration requirement imposes a practical minimum size of the survey area. Fig.~\ref{fig:zmSN}, taken at face value, implies that an efficient strategy for constraining $\mu$ with tSZ could be to make incredibly deep measurements on a single very massive and low-redshift cluster (or a handful of such clusters). If, however, the calibration for such a survey is to come from CMB anisotropy, the survey must contain enough sky in which the signal is dominated by primary CMB to achieve the required calibration precision. This disfavors strategies along the lines of pointing a powerful interferometer (such as ALMA) at a small number of massive clusters.

\section{Conclusion}
\label{sec:conclusion}

In this study, we have demonstrated that the spectrum of the tSZ effect in the direction of massive clusters of galaxies can be used to constrain the $\mu$-distortion monopole. 
We have shown that this can in principle be achieved without measuring the mean intensity across the sky and instead using a differential experiment that calibrates off of the CMB anisotropies, even when assuming the underlying CMB is an undistorted blackbody. 
We forecasted constraints on $\mu$ using the tSZ spectrum for the upcoming CMB-S4 experiment, using both the Wide and Deep surveys, as well as the proposed CMB-HD experiment. We found that the most massive clusters at the lowest redshifts provide the strongest constraints on the $\mu$-distortion monopole. In terms of raw sensitivity, we found that all three surveys closely match or outperform \cobefiras\ in constraining the $\mu$-distortion monopole. Extragalactic and galactic foregrounds significantly degrade these constraints to the point where CMB-S4 performs worse than \cobefiras, and CMB-HD delivers roughly equivalent constraints to \cobefiras. Specifically, we found that radio point sources heavily impact low-noise surveys such as S4-Deep and CMB-HD, whereas the inclusion of AME significantly degrades S4-Wide constraints on $\mu$. 

To improve on these constraints, foreground removal is a priority. Improving foreground removal, in general, requires additional frequency channels to help distinguish signal from foregrounds. In regards to specific foregrounds, improved masking of radio point sources using higher-resolution surveys should reduce their impact, while to reduce the effects of galactic foregrounds such as AME, the most straightforward strategy is to perform deep sky observations that avoid the galactic plane. We see in Fig.\ \ref{fig:muy} and Fig.\ \ref{fig:weights} that the distortion of the tSZ spectrum increases at lower frequencies. This suggests additional coverage at low-frequencies should also improve constraints on $\mu$. Finally, the inclusion of external data (particularly X-ray data) could be useful both in filling in the low-redshift gaps in the CMB experiments' cluster selection and in providing external priors on the temperature of individual clusters, helping to break degeneracies between $\mu$, central tSZ decrement, and cluster temperature.

Based on the above discussion, an experiment that would improve on these current constraints 
should have many frequency channels to remove foregrounds, with some channels dedicated to frequencies below 30 GHz if possible. The experiment should have beams comparable to the targeted cluster sample with white noise levels comparable to or better than CMB-S4. This implies observations with radio instruments combined with a CMB experiment have the potential to improve measurements of $\mu$. While our results suggest a deep observation of individual low-redshift clusters would be ideal for obtaining better constraints on $\mu$, we caution that one would need to also measure in the same observation the CMB at a high enough SNR for all frequencies to calibrate off of CMB anisotropy. 

Certain assumptions we make in our forecasts may turn out to be overly optimistic.  For example, we modeled each cluster as spherical and isothermal, which is not true of realistic clusters. Furthermore, the assumption of 100\% correlation between the foreground power across all frequency bands must break down at some level. Although the level of decorrelation in galactic dust at these frequencies has been limited to be very small \cite{Sheehy:2017gfx}, even a low level of decorrelation could degrade precision constraints significantly. 

The low-frequency enhancement of $\mu$-distortions of the tSZ effect
suggests that a synergistic combination of CMB and radio telescope data could further improve constraints on the $\mu$ monopole using this technique.   To realize this promise with specific radio surveys, future studies can use the  forecasting framework presented here to address the calibration and foreground-mitigation requirements of the combined data set.

\acknowledgments

We thank Andrey Kravtsov for useful conversations. D.Z.
was supported by the National Science Foundation Graduate Research Fellowship Program under Grant No. DGE1746045.  W.H. was supported by U.S.
Dept. of Energy contract DE-FG02-13ER41958 and the
Simons Foundation. T.C.~acknowledges support from National Science Foundation award OPP-1852617.
\appendix

\section{Flat Sky Harmonics}
\label{sec:flat}

For a function  on the sky $y(\vec{n}=\{\theta,\phi\})$ with support only on a small area $\theta\ll 1$ around the pole (in the main text, center of the cluster), we can directly relate the spherical 
harmonic $y_{\ell m}$ and flat sky $y_\ell$ harmonic coefficients
\begin{align}
y(\vec{n}) &= \sum_{\ell m} y_{\ell m} Y_{\ell m}(\vec{n}) \nonumber\\
& \approx \int \frac{ d^2 \ell}{(2\pi)^2} y(\vec \ell) e^{i\vec n\cdot \vec{\ell}},
\label{eq:expansion}
\end{align}
using an approximation for $Y_{\ell m}$ itself in an elaboration of the derivation in Ref.~\cite{Hu:2000ee}.   This approximation follows from the relation
(\cite{1994tisp.book.....G}, 8.722.2) 
\begin{equation}
\ell^m P_\ell^{-m} (\cos\theta) \approx J_m(\ell \theta)
\label{eq:PlmJm}
\end{equation}
for $m\ge 0$ and $\ell\gg 1$.   We can use the fact that 
\begin{eqnarray}
J_{-m}(x) &=& (-1)^m J_m(x), \\
P_\ell^{-m} &=& (-1)^m \frac{ (\ell-m)!}{(\ell+m)!} P_\ell^m,
\end{eqnarray} 
and
\begin{equation}
Y_{\ell m} = \sqrt{ \frac{2\ell+1}{4\pi} \frac{(\ell-m)!}{(\ell+m)!} }P_\ell^m(\cos\theta)e^{i m\phi}
\end{equation}
to obtain for all $m$
\begin{equation}
Y_{\ell m} \approx 
\ell^{-|m|} \sqrt{\frac{(\ell+|m|)!}{(\ell-|m|)!} }(-1)^m  \sqrt{ \frac{2\ell+1}{4\pi} }J_{m}(\ell\theta)e^{i m\phi}.
\end{equation}
When transforming functions with support only near the pole only $|m| \ll \ell$ modes contribute substantially due to the rapid variation of higher modes with $\phi$, so it is a good approximation to cancel the factorials with $\ell^{-|m|}$ and use
\begin{equation}
Y_{\ell m} \approx (-1)^m  \sqrt{ \frac{2\ell+1}{4\pi} }J_{m}(\ell\theta)e^{i m\phi}, \quad (|m| \ll \ell).
\end{equation}
  Note that we can always orient the pole of the spherical coordinate system to align with the region of support. 
We can now obtain the desired relation between the two coefficients in Eq.~(\ref{eq:expansion})  \cite{Hu:2000ee}
\begin{eqnarray}
y(\vec \ell) & \approx & \sqrt{\frac{4\pi}{2\ell+1} } \sum_m i^{-m} y_{\ell m} e^{i m\varphi_\ell} ,\nonumber\\
y_{\ell m} & \approx & \sqrt{\frac{2\ell+1}{4\pi} }\int  \frac{d\varphi_\ell}{2\pi} e^{-i m\varphi_\ell } y(\vec{\ell}),
\end{eqnarray}
where $\varphi_\ell$ is the azimuthal angle $\phi$  that $\vec\ell$ points at the pole.  In particular if the function is azimuthally symmetric around the pole only $m=0$ coefficients contribute and 
\begin{eqnarray}
y(\vec{\ell}) 
=  2\pi \int \theta d\theta J_0(\ell \theta)  y(\theta)
\approx \sqrt{\frac{4\pi}{2\ell +1} }y_{\ell 0} .
\end{eqnarray} 
It is common in the literature to slightly improve on the accuracy of the underlying approximation (\ref{eq:PlmJm}) at low $\ell$ by taking the argument of the Bessel function as $\ell \theta \rightarrow (\ell + 1/2)\theta$ and correspondingly $\ell^2 \rightarrow \ell(\ell+1)$, e.g.\ in the Gaussian beam profile formula (\ref{eq:beam}).

\section{Relativistic Corrections}
\label{sec:relativistic}

Following  Refs.~\cite{Challinor:1997fy,Itoh:1997ks}, we can derive the relativistic corrections to the $y$ distortion of an initial $\mu$ distortion using the generalized Kompaneets equation which is the expansion of the Compton collision term to the Boltzmann equation in the small energy transfer due to scattering.
  To first order in $\theta_e \equiv k_B T_e/m_e c^2$, 
  Eq.~(\ref{eq:Kompaneets}) is generalized to
\begin{align}
 \frac{\partial f}{\partial \tau} =  \theta_e \sum_{n=1}^4 x_e^n I_n \left[(1+f)\left(\frac{\partial}{\partial x_e} +1\right)^n  -  \frac{\partial^n\! f}{\partial x_e^n}  \right] f
 \label{eq:genKompaneets}
\end{align}
with
\begin{eqnarray}
I_1 & =&   4- x_e + (10- \tfrac{47}{2} x_e + \tfrac{21}{5} x_e^2 )\theta_e , \nonumber\\
I_2 &=&  1+  ( \tfrac{47}{2} - \tfrac{63}{5} x_e + \tfrac{7}{10} x_e^2 ) \theta_e ,\nonumber\\
I_3 &=&  ( \tfrac{42}{5} -\tfrac{7}{5} x_e )\theta_e, \nonumber\\
I_4 &=& \tfrac{7}{10} \theta_e .
\end{eqnarray}
We can again find the change $\Delta f$ in the $|y|\ll 1$ regime by plugging in an initial Bose-Einstein distribution to the right hand side of Eq.~(\ref{eq:genKompaneets}) to obtain
$
\Delta f = y x e^{x+\mu} f^2  g,
$
where
\begin{align} 
g =&    X -4 + \theta_e \Big[-10 + \frac{47}{2}   X - \frac{42}{5}   X^2 +\frac{7}{10}   X^3 
\nonumber\\
& +   S^2 (-\frac{21}{5} + \frac{7}{5}   X) + \frac{7 x^2}{10}(6-  X) \frac{T}{T_e}  \Big]
\label{eq:ggen}
\end{align}
and
\begin{equation}
  X = x \coth[ (x+\mu)/2], \quad   S = x \csch[ (x+\mu)/2].\label{eq:XS}
\end{equation}
Recall that the Comptonization parameter $y$ was defined in Eq.~(\ref{eq:y}) to vanish for $T=T_e$.

For the tSZ effect in clusters where $T_e\gg T$, the expression becomes even simpler, reproducing and generalizing the $\mu=0$ result found in Ref.~\cite{Itoh:1997ks}.
This same rule for generalizing $g$ in the presence of $\mu$ through the modification to $X$ and $S$ in Eq.~(\ref{eq:XS}) applies to the higher order in $\theta_e$ terms of  Ref.~\cite{Itoh:1997ks} for $T_e\gg T$ as we have explicitly checked to 4th order.
\vfill
\pagebreak

\bibliography{main}
\vfill
\end{document}